\documentclass[conference]{IEEEtran}
 
\usepackage{graphicx}
\usepackage{latexsym}
\usepackage{amsfonts}
\usepackage{amssymb}
\usepackage{psfrag}
\usepackage{cite}
\usepackage{subfigure}
\usepackage{comment,cancel,float}
\usepackage{amsmath,epsfig,epstopdf,color,lineno,makeidx,booktabs,multirow}
\usepackage[normalem]{ulem}
\usepackage{stfloats}

\begin{document}

\title{Efficient Direct Detection of $M$-PAM  Sequences with Implicit CSI Acquisition for The FSO System}

\author{ \IEEEauthorblockN{Tianyu Song and Pooi-Yuen Kam}
\IEEEauthorblockA{Department of Electrical and Computer Engineering\\
National University of Singapore, Singapore 117583\\
Email: \{song.tianyu, elekampy\}@nus.edu.sg}
}

\maketitle

\begin{abstract}
Compared to on-off keying (OOK), $M$-ary pulse amplitude modulation ($M$-PAM, $M>2$) is more spectrally efficient. 
However, to detect $M$-PAM signals reliably, the requirement of accurate channel state information is more stringent. 
Previously, for OOK systems, we have developed a receiver that requires few pilot symbols and can jointly detect the data sequence and  estimate the unknown channel gain implicitly. 
In this paper, using the same approach, we extend our previous work and derive a generalized receiver for $M$-PAM systems. 
A Viterbi-type trellis-search algorithm coupled with a selective-store strategy is adopted, resulting in a low implementation complexity and a low memory requirement.  
Therefore, the receiver is efficient in terms of energy, spectra, implementation complexity and memory. 
Using theoretical analysis, we show that its error performance approaches that of maximum likelihood detection with perfect knowledge of the channel gain, as the observation window length increases.
Also, simulation results are presented to justify the theoretical analysis.
\end{abstract}


\IEEEpeerreviewmaketitle
\section{Introduction}

Free space optical (FSO) communications provide larger bandwidth with higher security and higher flexibility compared with conventional wireless communications. 
Due to the complexity of phase and frequency modulation, intensity modulation with direct detection (IM/DD) is used for most current FSO communication systems. 
To increase the system spectral efficiency, higher order modulation, such as $M$-ary pulse amplitude modulation ($M$-PAM), is preferred. 
Compared to FSO on-off keying (OOK) systems, which can be considered as binary-PAM systems, the $M$-PAM ($M>2$) system suffers more severely from signal intensity fluctuations due to   atmospheric turbulence and pointing errors. 
To recover the information reliably,  accurate channel state information(CSI), i.e., the instantaneous channel gain, is required.  
One way to acquire the CSI is to use pilot symbols, which results in a reduction of system spectral efficiency and thus is not so desirable. 
The receiver design problem is still challenged by the fact that the CSI is hard to acquire at the receiver side if no pilot symbols are used.

Based on the generalised likelihood ratio test (GLRT) principle, a maximum likelihood (ML) sequence detection (MLSD) receiver (the GLRT-MLSD receiver) is derived for OOK systems in \cite{Song2014Arobust}. 
In this paper, based on the same GLRT principle, we extend the GLRT-MLSD receiver to higher order PAM systems. 
This new GLRT-MLSD receiver requires very few pilot symbols for initialising.
After that, it can continuously estimate the instantaneous channel gain and adapt its decision metric accordingly without further insertion of pilot symbols. 
Since few pilot symbols are required and higher order PAM is adopted, the system is spectrally efficient.

To efficiently implement the GLRT-MLSD receiver, the Viterbi-type trellis-search algorithm and the selective-store strategy proposed in \cite{Song2014Arobust} are adopted. 
By adopting the Viterbi-type trellis-search algorithm, the search complexity is reduced to a very low level that is independent of the observation window length. 
With the help of selective-store strategy, we completely avoid potential decision ambiguities which can cause an error floor problem.   
By theoretical analysis and simulation, we have shown that as the observation window length increases, the bit error probability (BEP) of our receiver approaches the Genie Bound, which is defined as the BEP of ML detection with perfect CSI (PCSI).

The remaining parts of this paper are organized as follows.
In section II, we briefly present the channel model.
Section III introduces the ideal ML detection with PCSI and the Genie Bound. 
Our GLRT-MLSD receiver for high-level PAM systems is derived in section IV.
The Viterbi-type trellis-search algorithm as well as the selective-store strategy is given in section V; also, we analyze the performance of our receiver with this implementation method.
Numerical results are given in section VI and our final conclusions are drawn in section VII. 
Throughout the paper, common italic font letters, such as $r$, $m$ and $L$, are used to denote scalar quantities;
bold non-italic font lower-case letters, such as $\mathbf{r}$ and $\mathbf{m}$, are used to denote vectors.

\section{System Model}
\begin{figure}
\centering
\includegraphics[scale=0.9]{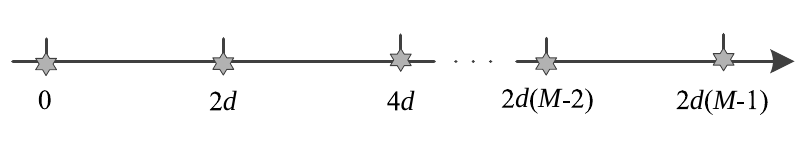}
\caption{Constellation at the transmitter side.}
\label{fg:Constellation_Tx}
 \end{figure}
We consider an IM/DD system using $M$-PAM with Gray coding, whose transmitter-side signal constellation is given in Fig. \ref{fg:Constellation_Tx}.   
The received electrical signal at time point $k$ is modelled as
\begin{align}
\label{eq:digitalchannel}
r(k)=2dhm(k)+n(k),
\end{align}
where  $2d$ is the minimum signal distance when no fading exists and is expressed in terms of the electrical-domain energy per bit $E_b$ and the modulation order $M$ by  
\begin{align}
2d = \sqrt{\frac{6E_b \log_2 M }{(M-1)(2M-1)}}.
\label{eq:A_single}
\end{align} 
In \eqref{eq:digitalchannel}, $h$ is the instantaneous channel gain,  $m(k)$  is the transmitted symbol which takes on any value from set $\{0, 1, ... ,M-1\}$ with equal probability, and  $n(k) $ is the additive white Gaussian noise (AWGN) with mean zero and power spectral density (PSD) $\frac{N_0}{2}$, i.e., $\mathbb{E}[n(i)n(j)]=\frac{N_0}{2} \delta _{ij}$.

As discussed in   \cite{Song2014Arobust} and \cite{Farid2007Outage}, geometric spread and pointing errors $h_p$, atmospheric turbulence $h_a$, and path loss $h_l$ together determine the overall channel gain $h$. The channel state $h$ can be formulated as  $h=h_p h_a h_l $, where $h_l$ is deterministic while $h_p$ and $h_a$ are stochastic. 
Therefore, without loss of generality, $h_l$ can be incorporated into $h_a$ which amounts to setting $h_l=1$, resulting to $h=h_{a} h_p$.
For the weak turbulence channel, $h_a$ is a log-normal distributed random variable with its probability density function (pdf):   
\begin{align}
p_{h_a}(h)=\frac{1}{2h\sigma_X \sqrt{2\pi}}\exp \left( -\frac{(\ln h)^2}{8\sigma_X^2}\right),\ h>0 ,
\label{eq:lndist}
\end{align}
where $\sigma_X^2$ is the variance of log-amplitude fluctuation. 
For the strong turbulence channel, $h_a$ is modelled by the Gamma-Gamma distribution with its pdf:  
\begin{align}
\label{eq:ggdist}
p_{h_a}(h)=\frac{2(\alpha\beta)^{(\alpha+\beta)/2}}{\Gamma(\alpha)\Gamma(\beta)}
h^{(\alpha+\beta)/2-1}
K_{\alpha-\beta}  \Big( 2\sqrt{\alpha\beta h} & \Big),   \\     & h>0 ,  \nonumber  
\end{align} 
where $K_{a}(\cdot)$ is the modified Bessel function of the second kind, and $1/\beta$ and $1/\alpha$ are the variances of the small and large scale eddies, respectively. 
The pdf of $h_p$  is expressed as
\begin{align}
p_{h_p}(h)=\frac{\gamma^2}{A_0^{\gamma^2}}h^{\gamma^2-1}, \quad  0<h<A_0,
\label{eq:pnte}
\end{align}  
where $A_0$ is the fraction of the collected power when no pointing error occurs, and $\gamma$  is the ratio between the equivalent beam radius at the receiver and the pointing error displacement standard deviation at the receiver.
Finally, the pdf of  $h=h_p h_a h_l $ can be derived by
\begin{align}
p_h(h)=\int_{0}^\infty \frac{1}{|a|}  p_{h_a}(a)p_{h_p}\left(\frac{h}{a}\right) da,\quad  h>0 .
\label{eq:h_pdf}
\end{align}
Further details of this channel model, especially    \eqref{eq:lndist} - \eqref{eq:pnte}, can be found in \cite{Zhu2002FSOComm, Farid2007Outage, karp1988optical, Andrews2001Mathe, Deva2009PointingE}.

Scintillation index (SI) is defined as the normalised variance of the irradiance fluctuations \cite{Mudge2011SI}. 
According to our model $h=h_p h_a h_l $, SI is given by
$\text{SI}= {\mathbb{E}(h_a^2 )}/{\mathbb{E}^2(h_a)} -1$,
but not $\mathbb{E}(h^2)/\mathbb{E}^2(h)-1$.

Since the time scales of these fading processes are of the order of  $10^{-3}$s to $10^{-2}$s, which is far larger than the bit interval ($\approx 10^{-9}$s for multi-Gbps systems), $h$ is considered to be constant over a large number $L_c$ of transmitted data symbols \cite{tunick2007statistical, Xie2011EffICC}.
It is obvious that the channel coherence length $L_c$ could be assumed at least of the order of $10^4$ symbol intervals.

The received electrical instantaneous signal-to-noise ratio (SNR) per bit is defined as   
$\mathbf{SNR} = \frac{h^2E_b}{N_0}$ \cite{Farid2007Outage}.
The \emph{average SNR} over all possible channel states is thus $
\mathbf{SNR}_\mathrm{ave} =  \frac{\mathbb{E}(h^2)E_b}{N_0}$.
Specifically, if pilot symbols are required, for example, there are $P$ pilot symbols and $D$ data symbols in every data packet, the \emph{effective average SNR} is $ \mathbf{SNR}_\mathrm{ave}^e = \frac{(P+D)\mathbb{E}(h^2)E_b}{D N_0}$.
If $P$ is very small compared to $D$, i.e., the ratio $\frac{P}{D}$ is far smaller than 1, the value of $\mathbf{SNR}_\mathrm{ave}^e$ tends to be equal to the value of $\mathbf{SNR}_\mathrm{ave}$.

In this paper, when it comes to pilot symbols, we specifically refer to the pilot symbols that are used for instantaneous CSI acquisition. Thus, $P$ is the number of pilot symbols that are used for CSI acquisition in each packet. In practice, pilot symbols are also required for some other purposes such as timing synchronization.

\section{The PCSI Receiver and the Genie Bound}

\begin{figure}
\centering
\includegraphics[scale=0.9]{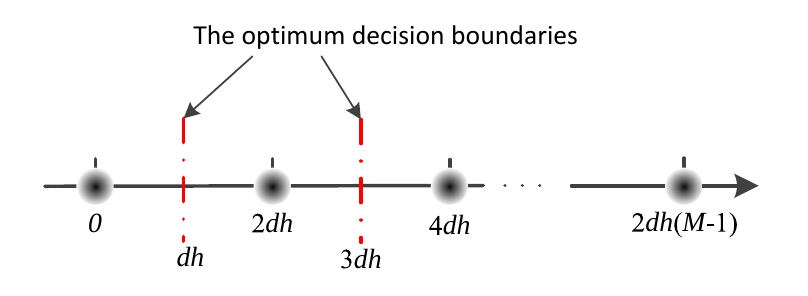}
\caption{Constellation at the receiver side}
 \label{fg:Constellation_Rx}
\end{figure}

The PCSI receiver knows the CSI perfectly and performs ML detection by setting the decision $\hat m(k)$ as 
\begin{align}
\hat{m}(k) = \arg \max_{m(k)} \  p(r(k)|h,d,m(k)),
\label{eq:rule0_pcsi}
\end{align}
where  $p(r(k)|h,d,m(k))$is the pdf of $r(k)$ conditioned on $h$, $d$ and $m(k)$.
Based on  \eqref{eq:digitalchannel}, we know this conditional pdf is 
\begin{align}
p(r(k)|h,d,m(k)) = \frac{1}{\sqrt{\pi N_0}} \exp \left(-\frac{(r(k) - 2dhm(k))^2}{N_0}  \right).
\label{eq:r_pdf}
\end{align} 
Substituting   \eqref{eq:r_pdf} into   \eqref{eq:rule0_pcsi} and eliminating irrelevant items, we simplify the decision rule to
\begin{equation}
\hat{m}(k) =  \arg \min_{m(k)} (r(k) - 2dhm(k))^2.
\label{eq:pcsi_rule0}
\end{equation}
After performing \eqref{eq:pcsi_rule0}, a reverse Gary mapping is required to recover the information bits.

At this point, we see that besides the CSI, i.e. the instantaneous value of $h$, the value of $d$ (or the value of $E_b$) is also essential to perform ML detection.
More intuitively, we give the constellation at the receiver side in Fig. \ref{fg:Constellation_Rx}, from which we can see the instantaneous values of both $h$ are $d$ are required.

{ 
We first consider the AWGN channel, in which the value of $h$ is always 1. 
The symbol error probability (SEP) of PAM signals over the AWGN channel is \cite[Eq. (4.3-4)]{proakis2008digital}
\begin{align}
P_s^\mathrm{PAM} (\frac{(2d)^2}{N_0},M)  = 
 \frac{2(M-1)}{M}   Q\left(\sqrt{\frac{2d^2}{N_0}}\right),
\label{eq:sep_pam}
\end{align}
where the $Q$-function $Q(\cdot)$ is defined as 
\begin{align}
Q(x) = \int^\infty_x \frac{1}{\sqrt{2\pi}}\exp\left(-\frac{x^2}{2}\right) dx. \nonumber 
\end{align}
Thus, the SEP of the PCSI receiver given in \eqref{eq:pcsi_rule0} conditioned on a fixed value of $h$ is $P_s^\mathrm{PAM} (\frac{(2dh)^2}{N_0},M)$ and 
the average SEP over all possible values of $h$ is 
\begin{align}
P_s^\mathrm{PCSI}(e) = \int_0^\infty P_s^\mathrm{PAM}(\frac{(2dh)^2}{N_0},M) p_h(h)dh .
\label{eq:sep_PCSI}
\end{align}

If only adjacent symbol errors are considered, the BEP of PAM signals can be approximated in a simple form, which is 
\begin{align}
P_b^\mathrm{PCSI}(e) \approx 
 \frac{1}{\log_2M}  P_s^\mathrm{PCSI}(e)
\label{eq:BEP_pam_approxi}
\end{align}
for the FSO channel \cite{steve1999Amulti} \cite[Sec. 3.3]{hranilovic2005wireless}.

This ideal receiver and its performance are used as a benchmark when analysing other receivers.
The average BEP of this PCSI receiver is also referred to as the \emph{Genie Bound}. 
As CSI is known at the receiver side, no pilot symbols are required, i.e., $P=0$.

Apart from this subsection, in all the other parts of this paper, when it comes to the BEP, we refer to the average BEP over all possible channel states.
}

\section{The GLRT-MLSD Receiver}

We consider a subsequence with $L$ immediate past symbols of the entire sequence, where $L$ is much smaller than the channel coherence length $L_c$. 
At time $k$, the transmitted data subsequence is denoted by $\mathbf{m}(k,L)=[m(k-L+1), ... , m(k)]$ where  $m(i)\in\{0, 1, ... , M-1\}$  for any integer $i$. 
Similarly,  $\mathbf{r}(k,L)=[r(k-L+1),  ... , r(k)]$ and $\mathbf{n}(k,L)=[n(k-L+1),  ... , n(k)]$ are used to denote the corresponding received signal subsequence and noise subsequence. 
For simplicity of notation, we drop the index terms $k$ and $L$, and denote these quantities as $\mathbf{m}$, $\mathbf{r}$ and $\mathbf{n}$.
As $L$ is much smaller than $L_c$, the received subsequence could be modelled as 
\begin{align}
\mathbf{r}=2dh\mathbf{m}+\mathbf{n}.
\label{eq:seq_channel}
\end{align}
Due to the independence of  the fading gains and the AWGN, the conditional pdf of $\mathbf r$ is
\begin{equation}
p(\mathbf{r}|\mathbf{m},h)=\frac{1}{(\pi N_0)^{L/2}}\exp \left(-\frac{\|\mathbf{r}-2dh\mathbf{m}\|^2}{N_0}\right).
\label{eq:seqpdf}
\end{equation} 
Our GLRT-MLSD receiver jointly decides on $\mathbf{m}$ and $ h$ that maximize $p(\mathbf{r}|\mathbf{m}, h)$ \cite{van2004detection}.
We use $\mathbf{\hat m}$ and $\hat h$ to denote the detection result on subsequence $\mathbf{m}$ and the estimated value of $h$, respectively.
From  \eqref{eq:seqpdf}, the decision rule is reduced to
\begin{align}
 (\hat{\mathbf{m}},\hat{h})=\arg \underset{\mathbf{m},h} \min \|\mathbf{r}-2dh \mathbf{m}\|^2 .
 \label{eq:rule1} 
\end{align}
For a given $\mathbf{m}$, by differentiating $\|\mathbf{r}-2dh \mathbf{m}\|^2 $ with respect to $h$ and letting the derivative be equal to zero, we find that $\|\mathbf{r}-2dh \mathbf{m}\|^2 $  achieves its minimum value at
\begin{equation}
\hat h(\mathbf m)=\frac{\mathbf{r}\cdot\mathbf{m}}{2d\|\mathbf{m}\|^2} .
\label{eq:h_estimate}
\end{equation}
Here, it should be emphasized that $\hat h(\mathbf m)$ is the ML estimate of $h$ conditioned on a hypothesised data subsequence $\mathbf{m}$. 
Substituting   \eqref{eq:h_estimate} into $\|\mathbf{r}-2dh \mathbf{m}\|^2 $ , we get
\begin{equation}
 \|\mathbf{r}-2dh\mathbf{m}\|^2 \Big|_{h =\hat h(\mathbf m) } = \|\mathbf{r}\|^2 -\frac{(\mathbf{r}\cdot\mathbf{m})^2}{\|\mathbf{m}\|^2}.
 \label{extendsub}
\end{equation}
The right hand side of   \eqref{extendsub} gives the minimum value of $\|\mathbf{r}-2dh \mathbf{m}\|^2 $ for each subsequence $\mathbf{m}$. 
Thus, to minimize $\|\mathbf{r}-2dh \mathbf{m}\|^2 $, we only need to maximize $\frac{(\mathbf{r}\cdot\mathbf{m})^2}{\|\mathbf{m}\|^2}$ with respect to $\mathbf{m}$. 
The decision rule    \eqref{eq:rule1} now is reduced to
\begin{equation}
\hat{\mathbf{m}} =\arg \underset{\mathbf{m}} \max \ \lambda(\mathbf m) =  \arg \underset{\mathbf{m}} \max \ \frac{(\mathbf{r}\cdot\mathbf{m})^2}{\|\mathbf{m}\|^2}.
\label{eq:GLRT_metric}
\end{equation}
After   \eqref{eq:GLRT_metric}, a reverse Gray mapping, which is from symbols to bits, is required to recover the transmitted data information in bits.

Here, we can see that the decision metric of our GLRT-MLSD receiver given in   \eqref{eq:GLRT_metric} is exactly the same as the one given in \cite[Eq. (39)]{Song2014Arobust}. 
Thus, it retains all the benefits:
\begin{itemize}
\item Only a simple evaluation of the decision metric value, where no integrals  and no requirement of values of $E_b$ and $N_0$ are involved, is required
\item The estimation of $h$ in  \eqref{eq:h_estimate} is implicit
\item That the channel model information is not required in   \eqref{eq:GLRT_metric} makes it robust and enables it to operate in any slowly time-varying environment, regardless of the distribution of $h$
\end{itemize}

\section{The Suboptimal, Efficient Implementation}

\subsection{The Viterbi-Type Trellis-Search Algorithm}

\begin{figure}
\centering
\includegraphics[scale=1]{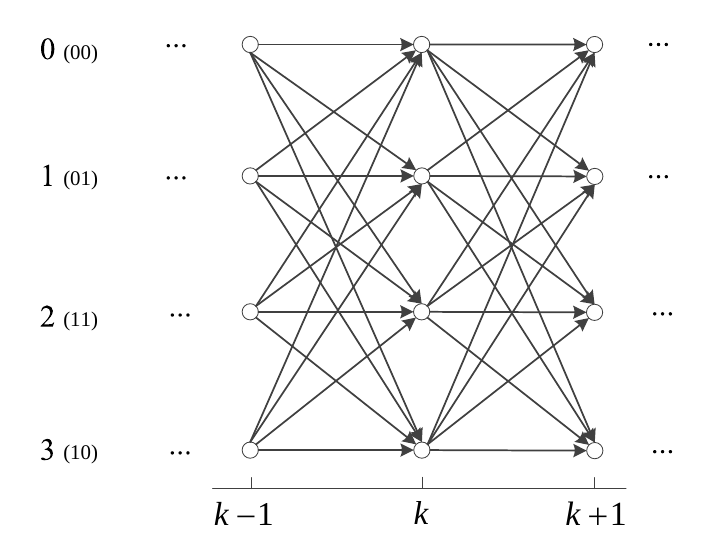}
\caption{Trellis Diagram of 4-PAM signals}
 \label{fg:4pam_trellis}
\end{figure}

In principle, to implement \eqref{eq:GLRT_metric}, one has to compare the metric values of $M^L$ possible subsequences and choose the one with highest metric value. 
Clearly, the search complexity increases exponentially with $L$. 
Thus, to use a large value of $L$ and to keep the implementation simple, a Viterbi-type trellis-search algorithm, which has been first investigated in \cite{kam1995viterbi} and further studied for FSO OOK systems in \cite{Song2014Arobust}, is adopted. 
OOK can be regarded as a special case of $M$-PAM for $M=2$.
In \cite[Fig. 2]{Song2014Arobust}, the trellis for OOK signals is given.
Here, we only give the trellis for 4-PAM in Fig. \ref{fg:4pam_trellis} and trellises with other values of $M$ can be obtained accordingly.
For generalised $M$-PAM signals, the search algorithms are similar and we provide an example for 4-PAM systems as follows.

As shown in Fig. \ref{fg:4pam_trellis}, a trellis with 4 nodes at each time point $k$ is constructed. 
Each node is labelled by the possible transmitted data symbol 0, 1, 2 or 3.  
All the branches, leading into the same node, represent transmission of the symbol corresponding to that node at time point $k$. 
The metric of the path $\mathbf m(k,L)$ at time $k$ is computed based on \eqref{eq:GLRT_metric} with only the $L$ most recent data symbols and received signals.  
For each node at each time $k$, four paths corresponding to four hypothesized subsequences enter it.   
The algorithm computes the metrics of the four hypothesized subsequences, then saves the one with the highest metric value, and discards others. 
The same is repeated for all paths entering the same node, and the path with the largest metric is saved as the survivor. 
The decision on a symbol is made only when the tails of all survivors have merged at a node. 
An additional reverse Gray mapping needs to be performed to recover the information bits.  

We segment the subsequence into two parts: the \emph{detected part} that is the subsequence before the most recent merge node, and the \emph{ongoing part} that is the part after the most recent node.  
We use $L_w$ to denote the length of the detected part and $l$ the ongoing part. 
The detected part is also called the observation window, and thus $L_w$ is the observation window length.  
In practice, to choose an appropriate memory size, three aspects need to be considered. 
One is to retain complete survivors after the most recent merge node. 
Obviously, $l$ is a random number. 
From simulation observations, the mean of $l$ is no larger than 3.
We can safely set $l=20$ to reserve some memory for possible long survivors. 
Another aspect is to choose a large $L_w$ to keep the probability that all the data symbols in the observation window are zeros as small as possible. 
If only zeros exist in the observation window, we will have a zero denominator of the decision metric \eqref{eq:GLRT_metric} and a consequent decision ambiguity. 
The last aspect is to make sure $L_w \ll L_c$.

\subsection{The Selective-Store Strategy} 

We just argued that it is possible to encounter a large number of successive zeros in the data stream filling the whole observation window, which would lead to a zero denominator and cause a consequent error floor problem. 
Increasing the value of $L_w$ can only reduce the probability of incurring a zero denominator but can not avoid it completely.  
Similar to that introduced in \cite{Song2014Arobust},  when storing the detected subsequence and the corresponding received signal subsequence element by element, we can discard element $\hat m (i) = 0 $ and the corresponding element $r(i)$ for any $i$. 
The memory size for the detected signals, i.e. the observation window size $L_w$, does not change, while we completely avoid the potential troublesome zero denominator. 

\subsection{Performance Analysis}

Since the decision metric of a path (corresponding to a subsequence) given in \eqref{eq:GLRT_metric} is not the sum of all branch metrics, the optimality of the detector output cannot be guaranteed.  
In this section, we prove that the BEP of this suboptimal detection method can approach the Genie Bound as the observation window length increases.

We suppose that the actual transmitted data subsequence at time $k+1$ is $\mathbf{m}_0(k+1,L') = [ m_0(k-L'+2), ... , m_0(k+1)]$, and $\mathbf{m}_1(k+1,L') = [m_1(k-L'+2) , ... , m_1(k+1)]$ is an alternative subsequence, where $m_0(i)$ and $m_1(i)$ can take any value from set $\{0, 1, ..., M-1\}$ with equal probability.   
Subsequence $\mathbf{m}_0$ differs from $\mathbf{m}_1$ only at time $k$; i.e., ${\mathbf{m}}_0(k-1,L'-2) = {\mathbf{m}}_1(k-1,L'-2)$,  $m_0(k+1) = m_1(k+1)$, and $m_0(k) \neq m_1(k)$.   
We further assume that exact $L_w$ non-zero data symbols exist in  subsequence $ {\mathbf{m}}_0(k-1,L'-2)$ .
Vector $\mathbf{r}(k+1, L')=2dh\mathbf{m}_0(k+1, L')+\mathbf{n}(k+1, L')$ is the received signal subsequence, where  $\mathbf{n}(k+1, L')= [n(k-L') , ... , n(k+1)]$ is the noise subsequence.
For simplicity of notation, in this subsection, we drop the index terms $k+1$ and $L'$, and use $\mathbf{m}_0$, $\mathbf{m}_1$ $\mathbf{r} $  and $\mathbf{n} $  to denote the  subsequences.
We know that paths of $\mathbf{m}_0$ and $\mathbf{m}_1$ merge at time $k-1$; i.e., the decisions before time $k-1$ (including $k-1$) have been made.
From the decision rule \eqref{eq:GLRT_metric}, the receiver will decide $\mathbf{\hat m} = \mathbf{m}_1$ only if the decision metric of $\mathbf{m}_1$ is higher than that of $\mathbf{m}_0$. 
The pairwise error probability, i.e., the probability of the event that the receiver decides in favor of $\mathbf{m}_1$ given that $\mathbf{m}_0$ is transmitted and $\mathbf{m}_1$ is the only other alternative sequence, is given by
\begin{align}
&P(\varepsilon |\mathbf{m}=\mathbf{m}_0,h)  \nonumber \\
=&P \left( \frac{(\mathbf{r}\cdot\mathbf{m_1})^2}{\|\mathbf{m_1}\|^2} > \frac{(\mathbf{r}\cdot\mathbf{m_0})^2}{\|\mathbf{m_0}\|^2}\bigg|\mathbf{m}=\mathbf{m_0},h \right) \nonumber \\
=&P\bigg( (\mathbf{m}_+\cdot \mathbf{r})(\mathbf{m}_- \cdot \mathbf{r}) >0
\bigg| \mathbf{m}=\mathbf{m}_0,h \bigg),
\end{align}
where $\mathbf{m}_+=\|\mathbf{m}_0\|\mathbf{m_1} 
+ \|\mathbf{m}_1\|\mathbf{m_0}$, and
$\mathbf{m}_-=\|\mathbf{m}_0\|\mathbf{m_1} 
- \|\mathbf{m}_1\|\mathbf{m_0}$.
We use $X_+$ to denote $\mathbf{m}_+\cdot\mathbf{r}$, and $X_-$ to denote $\mathbf{m}_-\cdot\mathbf{r}$. By the law of total probability, we have 
\begin{align}
&P(\varepsilon |\mathbf{m}=\mathbf{m}_0,h) 
= P(X_+X_->0|h) \nonumber \\
=& P(X_->0| X_+>0,h)P(X_+>0|h)   +  \nonumber \\
& \qquad \qquad \qquad  P(X_-<0| X_+<0,h)P(X_+<0|h)
\label{eq:PEP_total}
\end{align}

We first define $S$ as
\begin{align}
S = \left( \sum_{i=k-L'+2}^{k-1} m^2(i) \right) + m^2(k+1).
\end{align} 
Obviously, we have $L_w \leq S \leq (M-1)^2(L_w+1) $, $ \|\mathbf{m}_0\|^2 = S +m_0^2(k)$, $\|\mathbf{m}_1\|^2 = S +m_1^2(k)$ and $ \mathbf{m}_0\mathbf{m}_1  = S +m_0(k) m_1(k)$. 
We then examine the statistics of $X_+$ and $X_-$.
Since $\mathbf{r} =2dh\mathbf{m}_0 +\mathbf{n}$ and the components of the noise vector $\mathbf{n}$ are independently identically Gaussian distributed with mean zero and variance $N_0/2$,  $X_+=\mathbf{m}_+\mathbf{r}$ and $X_-=\mathbf{m}_-\mathbf{r}$ are Gaussian random variables. 
For $X_+$,  we have its mean 
\begin{align}
&\mathbb{E}[X_+]=\mathbb{E}[   \mathbf{m}_+\cdot\mathbf{r} ] 
=\mathbb{E}[   \mathbf{m}_+ \cdot(2dh \mathbf{m}_0 +\mathbf{n}) ]  \nonumber \\
=&2dh  (\mathbf{m}_+ \cdot  \mathbf{m}_0) 
= 2dh \|\mathbf{m}_0\|\mathbf{m}_1 \cdot \mathbf{m}_0 + 2dh\|\mathbf{m}_0\|^2 \|\mathbf{m}_1\|  \nonumber  \\ 
%=& 2dh \sqrt{S+m_0^2(k)} \times \nonumber \\ 
%&\left(S+m_0(k)m_1(k) +\sqrt{(S+m_0^2(k))(S+m_1^2(k)) }   \right)\nonumber  \\ 
=& \mu 2dh \sqrt{S+m_0^2(k)} 
\label{eq:e_X+}
\end{align}
and   variance  
\begin{align}
&\mathrm{Var}[X_+]=\mathrm{Var}[ \mathbf{m}_+\cdot\mathbf{r}]  
=\mathrm{Var} [ \mathbf{m}_+ \cdot(2dh \mathbf{m}_0 +\mathbf{n})]    \nonumber \\
=&\mathrm{Var} [ \mathbf{m}_+ \cdot \mathbf{n}] 
= \frac{\|\mathbf{m}_+\|^2N_0}{2}   
 = \frac{ \Big\|\|\mathbf{m}_0\|\mathbf{m_1} 
+ \|\mathbf{m}_1\|\mathbf{m_0}\Big\|^2 N_0}{2}  \nonumber \\  
% = &    N_0 \|\mathbf{m}_0\|^2 \|\mathbf{m_1}\|^2 + N_0
% \|\mathbf{m}_1\|\|\mathbf{m_0}\|    \mathbf{m}_1 \mathbf{m_0}    \nonumber \\
% =&  N_0(S+m_0^2(k))(S+m_1^2(k)) + \nonumber \\ 
% & N_0 \sqrt{S+m_0^2(k)}\sqrt{S+m_1^2(k)}(S+m_0(k)m_1(k))   \nonumber \\ 
% =& N_0 \sqrt{(S+m_0^2(k))(S+m_1^2(k))} \times  \nonumber \\
% & \Big( S+m_0(k)m_1(k) + \sqrt{(S+m_0^2(k))(S+m_1^2(k)) } \Big) \nonumber \\ 
 =& \mu N_0 \sqrt{(S+m_0^2(k))(S+m_1^2(k))} .
 \label{eq:var_X+}
\end{align} 
Similarly, the mean and variance of $X_-$ are
\begin{align}
&\mathbb{E}[X_-]
= 2dh  (\mathbf{m}_- \cdot  \mathbf{m}_0)  \nonumber \\
 =& 2dh \|\mathbf{m}_0\|\mathbf{m}_1 \cdot \mathbf{m}_0 - 2dh\|\mathbf{m}_0\|^2 \|\mathbf{m}_1\|  \nonumber  \\ 
%=& 2dh \sqrt{S+m_0^2(k)} \times \nonumber \\ 
%&\left(S+m_0(k)m_1(k) - \sqrt{(S+m_0^2(k))(S+m_1^2(k)) }   \right) \nonumber \\
%=& \frac{-2dhS\sqrt{S+m_0^2(k)}(m_1(k)-m_0(k))^2}{S+m_0(k)m_1(k) + \sqrt{(S+m_0^2(k))(S+m_1^2(k)) }  }  \nonumber \\
=& \frac{-2dhS\sqrt{S+m_0^2(k)}(m_1(k)-m_0(k))^2}{\mu}
\label{eq:e_X-}
\end{align}
and
\begin{align}
&\mathrm{Var}[X_-]
= \frac{\|\mathbf{m}_-\|^2N_0}{2} =\frac{ \Big\|\|\mathbf{m}_0\|\mathbf{m_1} 
- \|\mathbf{m}_1\|\mathbf{m_0}\Big\|^2 N_0}{2}  \nonumber \\
%=&\mathrm{Var}[X_-]
% = N_0(S+m_0^2(k))(S+m_1^2(k)) - \nonumber \\ 
% & N_0 \sqrt{S+m_0^2(k)}\sqrt{S+m_1^2(k)}(S+m_0(k)m_1(k)) \nonumber \\ 
%  = & N_0 \sqrt{S+m_0^2(k)}\sqrt{S+m_1^2(k)} \times  \nonumber \\ 
% & \Big( \sqrt{S+m_0^2(k)}\sqrt{S+m_1^2(k)}-S - m_0(k)m_1(k) \Big) \nonumber \\
%  = &\frac{N_0 S\sqrt{(S+m_0^2(k))(S+m_1^2(k))} (m_0(k)-m_1(k))^2} {S + m_0(k)m_1(k)+\sqrt{(S+m_0^2(k))(S+m_1^2(k)) }}  \nonumber \\
    = &\frac{N_0 S\sqrt{(S+m_0^2(k))(S+m_1^2(k))} (m_0(k)-m_1(k))^2} {\mu}
    \label{eq:var_X-}
\end{align} 
respectively. For simplicity, we define 
\begin{align}
\mu = S + m_0(k)m_1(k)+\sqrt{(S+m_0^2(k))(S+m_1^2(k)) }
\label{eq:mu}
\end{align}
in \eqref{eq:e_X+}-\eqref{eq:var_X-}.

Therefore, the probability $P(X_+>0|h)$ and $P(X_->0|h)$ can be easily obtained as:
\begin{align}
\label{eq:X+>0}
&P(X_+>0|h)=1-
Q\left(   \frac{ \mathbb{E}[X_+] }{\sqrt{ \mathrm{Var}[X_+]   }} \right)  
\nonumber \\
=& 1-  Q\left( \frac{2dh}{N_0} \left[\frac{S+m_0^2(k)}{S+m_1^2(k)}\right]^{1/4} \mu^{1/2} \right) 
\end{align}
and
\begin{align}
&P(X_->0|h)= 
Q\left( - \frac{ \mathbb{E}[X_-] } {\sqrt{ \mathrm{Var}[X_-]} } \right)  \nonumber \\
=&  Q\left(  \frac{2dh|m_0-m_1|}{\sqrt{N_0}} \left[\frac{S+m_0^2(k)}{S+m_1^2(k)}\right]^{1/4}  \sqrt{\frac{S}{\mu}} \right) .
\label{eq:X->0}
\end{align}

From \eqref{eq:mu} and $L_w \leq S \leq (M-1)^2(L_w+1)$, we know
\begin{align}
\lim_{L_w\rightarrow \infty} \mu = \infty
\end{align}
and 
\begin{align}
\lim_{L_w\rightarrow \infty} \frac{S}{\mu} = \frac{1}{2}.
\end{align}
Hence, we have 
\begin{align}
\label{eq:X+>0limit}
\lim_{L_w\rightarrow \infty} P(X_+>0) &= 1, \\
\label{eq:X+<0limit}
\lim_{L_w\rightarrow \infty} P(X_+<0) &= \lim_{L\rightarrow \infty} [1-P(X_+>0)]= 0, 
\end{align}
and
\begin{align}
\label{eq:X->0limit}
 \lim_{L_w\rightarrow \infty} P(X_->0) = Q\left(\frac{2dh|m_0(k)-m_1(k)|}{\sqrt{2N_0}} \right).
\end{align}

From \eqref{eq:PEP_total}, \eqref{eq:X+>0limit}, \eqref{eq:X+<0limit}, \eqref{eq:X->0limit} and the   \cite[\emph{Lemma 1}]{Song2014Arobust}, we see that the conditional pairwise error probability of our GLRT-MLSD receiver when $L$ goes to infinity is  
\begin{align}
&\lim_{L_w\rightarrow \infty}P(\varepsilon |\mathbf{m}=\mathbf{m}_0,h)  \nonumber \\
=&\lim_{L_w\rightarrow \infty}[ P(X_->0| X_+>0,h)P(X_+>0|h)   +  \nonumber \\
& \qquad \qquad \qquad \qquad P(X_-<0| X_+<0,h)P(X_+<0|h)] \nonumber \\ 
=&\lim_{L_w\rightarrow \infty} P(X_->0|h)   
= Q\left(\frac{2dh|m_0(k)-m_1(k)|}{\sqrt{2N_0}} \right).
\label{eq:pe-limit}
\end{align}

The conditional pairwise error probability of the PCSI receiver is \cite[Eq. (4)]{UysalBER2004}
\begin{align}
 P_\mathrm{PCSI}(\varepsilon |\mathbf{m}=\mathbf{m_0},h)   
=  Q\left(\frac{2dh|m_0(k)-m_1(k)|}{\sqrt{2N_0}} \right) .
\label{eq:pe=pcsi}
\end{align}

Comparing \eqref{eq:pe-limit} and \eqref{eq:pe=pcsi}, we can draw our conclusion: if the observation window length $L_w$ is  sufficiently large, the error performance of our GLRT-MLSD receiver approaches that of the PCSI receiver for all values of the channel gain $h$.

\section{Numerical Results and Discussions}

\begin{figure}
\centering
\subfigure[SI = 0.0941.]{
\includegraphics[scale=.9]{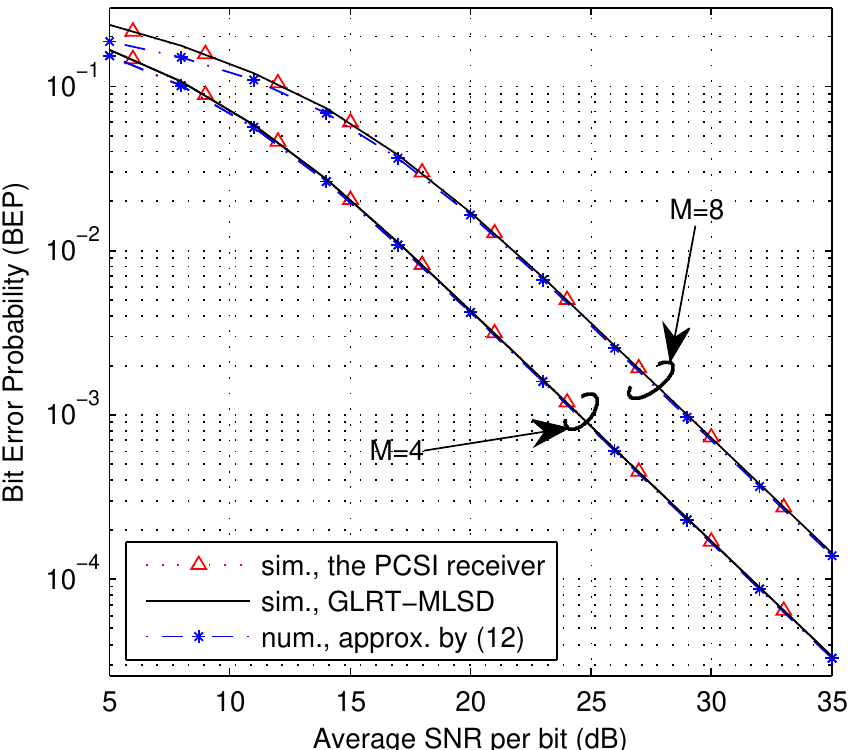}
\label{fg:BEP_w}
}
\centering
\subfigure[SI = 1.3890.]{
\includegraphics[scale=.9]{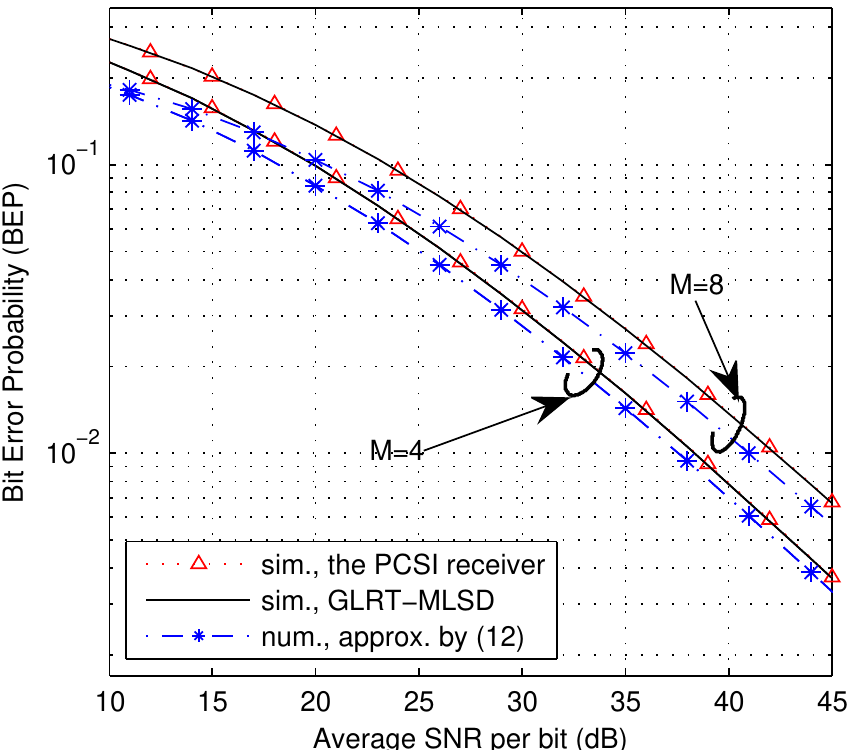}
\label{fg:BEP_s}
}
\caption{Performance over channels with different SI values. }
 \label{fg:BEP}
\end{figure} 
For the numerical results of this paper, we choose parameter $\sigma_X=0.15$ of the log-normal distribution for the weak turbulence channel, parameters $\alpha=2.23$ and $\beta=1.54$ of the Gamma-Gamma distribution for the strong turbulence channel, $A_0  = 0.0198$ and $\gamma^2 = 2.8071$ for the pointing errors. 
The corresponding SI values for the weak turbulence channel and the strong turbulence channel are 0.0941 and 1.3890, respectively. 
According to\mbox{\cite{andrews2001laserScin}}, the turbulence with SI=0.0941, which is less than 1, is in weak irradiance fluctuations regime; and the turbulence with SI=1.3890, which is larger than 1, is in moderate-to-strong irradiance fluctuations regime.

In Fig. \ref{fg:BEP}, we plot our numerical results.
The curve with a legend "sim." is obtained by simulation and the curve with a legend "num." is obtained by numerical integration.
Apparently, with $L_w=100$, the BEP of our GLRT-MLSD receiver perfectly achieves the Genie Bound for both SI = 0.0941 in Fig. \ref{fg:BEP_w} and SI = 1.3890 in  Fig. \ref{fg:BEP_s}.
This observation completely agree with the conclusion we just obtained by theoretical analysis.

We observe that the approximate BEP expression \eqref{eq:BEP_pam_approxi} is more accurate for lower-$M$-value cases.
This is because that in \eqref{eq:BEP_pam_approxi}, only adjacent symbol error is considered.
For lower-$M$-value cases, the total number of possible non-adjacent error symbols  is low. 
A typical example is that when $M=2$, i.e., for OOK systems, only adjacent symbol errors exist. 
We also observed that \eqref{eq:BEP_pam_approxi} is more accurate at lower BEP regions. 
This is because in lower BEP regions, non-adjacent symbol errors occur less frequently compared to the adjacent symbol error.

It should be noted that, though the search complexity of our Viterbi-type trellis-search algorithm is independent of $L_w$, it increases almost quadratically with  $M$.
More specifically, the search complexity per bit detection is $M^2/\log_2M$.
For a high value of $M$, the complexity of this GLRT-MLSD receiver is  very high. 
In our paper, we just give our simulation results when $M=4$ and $M=8$.

\section{Conclusions}
In this paper, we extend our previous work in \cite{Song2014Arobust} and derive a generalized Viterbi-type trellis-search sequence receiver that jointly detects the data sequence and estimates the unknown channel gain implicitly.
The decision metric of this receiver can be evaluated with a very low computational complexity.
It is robust in that it continuously performs ML channel estimation  without knowledge of the channel model, and adapts the decision metric accordingly.
Due to the adoption of the Viterbi-type trellis search algorithm, this receiver can be implemented with a low search complexity that is independent of the observation window length. 
With the help of the selective store strategy, we completely avoid the potential zero-metric-denominator problem which will cause an error floor. 
The BEP of our generalized receiver has been shown by both theoretical analysis and simulation to approach the Genie Bound, as the observation window increases.


\end{document}